\documentclass[11pt]{article}
\title{Particle-Like Solutions of the Einstein-Dirac-Maxwell Equations}
\author{Felix Finster\thanks{Research supported by the 
Deutsche Forschungsgemeinschaft and the Schweizerischer 
Nationalfonds.},\ 
Joel Smoller\thanks{Research supported in part by the NSF, Grant No.\ 
DMS-G-9501128.}, and Shing-Tung Yau\thanks{Research supported in part 
by the NSF, Grant No.\ 33-585-7510-2-30.}}
\date{February 1998}

\newcommand{\spc}{\;\;\;\;\;\;\;\;\;\;}

\newcommand{\R}{\mbox{\rm I \hspace{-.8 em} R}}

\begin{document}
\include{epsf}

\maketitle
\begin{abstract}
We consider the coupled Einstein-Dirac-Maxwell equations for a static, 
spherically symmetric system of two fermions in a singlet spinor 
state. Soliton-like solutions are constructed numerically.
The stability and the properties of the ground state solutions are
discussed for different values of the electromagnetic 
coupling constant. We find solutions even when the electromagnetic 
coupling is so strong that the total interaction is repulsive in the 
Newtonian limit. Our solutions are regular and well-behaved; this 
shows that the combined electromagnetic and gravitational 
self-interaction of the Dirac particles is finite.
\end{abstract}

\section{Introduction}
\setcounter{equation}{0}
In the recent paper \cite{FSY}, we studied the coupled Einstein-Dirac 
(ED) equations for a spherically symmetric system of two fermions in 
a singlet spinor state. Using numerical methods, we found particle-like
solutions and analyzed their properties. In this letter, we extend our
results to the more physically relevant system where the Einstein-Dirac
equations are coupled to an electromagnetic field. We find 
particle-like solutions which are linearly stable (with respect to 
spherically symmetric, time-dependent perturbations), and have other
interesting properties.

The general Einstein-Dirac-Maxwell (EDM) equations for a system of $n$ Dirac 
particles are
\begin{equation}
	R^i_j-\frac{1}{2} \:R\:\delta^i_j \;=\; -8 \pi \:T^i_j 
	\;\;\;,\;\;\;\;\;
	(G-m) \:\Psi_a \;=\; 0 \;\;\; , \;\;\;\;\;
	\nabla_k F^{jk} \;=\; 4 \pi e\sum_{a=1}^n \overline{\Psi_a} G^j \Psi_a
	\;\;\; , \label{1.1}
\end{equation}
where $T^i_j$ is the sum of the energy-momentum tensor of the Dirac
particles and the Maxwell stress-energy tensor.
Here the $G^j$ are the Dirac matrices, which are related to the Lorentzian
metric via the anti-commutation relations
\[ g^{jk}(x) \;=\; \frac{1}{2} \left\{ G^j(x),\: G^k(x) \right\} \;\;\; , \]
$F_{jk}$ is the electromagnetic field tensor, and $\Psi_a$
are the wave functions of fermions of mass $m$ and charge $e$.
The Dirac operator is denoted by $G$; it depends on both the 
gravitational and the electromagnetic field (for details, see e.g.\ \cite{F}).
In order to get a spherically symmetric system, we consider, as in \cite{FSY},
two fermions with opposite spin, i.e.\ a singlet spinor state.
Using the ansatz in \cite{FSY}, we reduce the Dirac $4$-spinors to a
2-component real spinor system $(\alpha, \beta)$.
We show numerically that the EDM equations also have particle-like 
solutions, which are characterized by the ``rotation number'' 
$n=0,1,\ldots$ of the vector $(\alpha, \beta)$. In contrast to 
\cite{FSY}, we restrict our attention to studying the case $n=0$, the 
ground state; this case illustrates quite nicely the 
physical effects due to the addition of the electromagnetic 
interaction. We anticipate that the situation for solutions
with higher rotation number will be qualitatively similar.

The relative coupling of the electromagnetic and the gravitational 
field is described by the parameter $(e/m)^2$. If we consider the 
ground state solutions for fixed value of $(e/m)^2$, we find that
the mass-energy spectrum (i.e., the plot of the binding energy of the
fermions vs.\ the rest mass)
is a spiral curve which tends to a limiting configuration $\Gamma$.
This implies the interesting result that for parameter values on
$\Gamma$, there exist an infinite number of $(n=0)$-solutions
(the one of lowest energy is the ground state), while for
parameter values near $\Gamma$, there are a large, but finite number of
such solutions. For small coupling (i.e., for small $m$ and $(e/m)^2<1$),
our solutions are linearly stable with respect to spherically
symmetric perturbations. If we compare the (ADM) mass $\rho$ with the 
total rest mass $2m$, we find that for small $m$, the total binding 
energy $\rho - 2m$ is negative, implying that energy is gained in 
forming the singlet state. It is thus physically reasonable that such 
states should be stable. However, the stable solutions become
unstable as the binding energy of the fermions increases; this
is shown using Conley index methods together with bifurcation theory (see 
\cite[Part IV]{S}).

In order to study the effect of the electromagnetic interaction in 
more detail, we look at the behavior of the solutions as the parameter
$(e/m)^2$ is varied.
For weak electromagnetic coupling $(e/m)^2 \ll 1$, the solutions
are well-behaved and look similar to the solutions of the ED equations
\cite{FSY}.
For $(e/m)^2 > 1$, the form of the solutions changes drastically.
In a simplified argument, this can already be understood from the 
nonrelativistic, classical limit of the EDM equations.
Namely, according to Newton's and Coulomb's laws, the force between two
charged, massive point particles has the well-known form
\begin{equation}
	F \;=\; -\frac{m^2}{r^2} \:+\:\frac{e^2}{r^2}
	\label{newton}
\end{equation}
(we work in standard units $\hbar=c={\cal{G}}=1$).
For $(e/m)^2<1$, the gravitational attraction dominates the 
electromagnetic repulsion, and it therefore seems
reasonable that we get bound states. For $(e/m)^2>1$, however,
the total force is repulsive, and clasically one can no longer expect
bound states. The EDM equations, however, do have solutions
even for $(e/m)^2>1$. This is a surprising effect 
which can again only be explained by the nonlinearity of Einstein's 
equations. For such solutions to exist, however, the rest mass $m$ of the
fermions must be sufficiently large.

We remark that there are related works \cite{C, B, L},
where the authors obtain soliton solutions for the Dirac-Maxwell
equations (in the absence of gravity), but these solutions,
unfortunately, have the undesirable feature of having negative
energy.

\section{The Equations}
\setcounter{equation}{0}
We choose polar coordinates $(t, r, \vartheta, \varphi)$ and write the 
metric in the form
\[ ds^2 \;=\; \frac{1}{T^2} \:dt^2 \:-\: \frac{1}{A} \:dr^2 \:-\: 
r^2 \:d \vartheta^2 \:-\: r^2 \:\sin^2 \vartheta \: d\varphi^2
\]
with positive functions $A(r)$, $T(r)$.
Using the ansatz from \cite[Eqns.\ (3.4),(3.6)]{FSY}, we describe the Dirac
spinors with two real functions $\alpha$, $\beta$. For the derivation of the
corresponding EDM equations, we simply modify the ED equations
\cite[Eqns.\ (5.4)-(5.8)]{FSY}. In the Dirac equation, the electromagnetic 
field is introduced by the minimal coupling procedure $\partial_j 
\rightarrow \partial_j - i e {\cal{A}}_j$, where $e$ is the unit charge and 
${\cal{A}}$ the electromagnetic potential (see \cite{BD}). In the 
static case, the fermions only generate an electric field; thus we 
can assume that the electromagnetic potential has the form 
${\cal{A}}=(-\phi, \vec{0})$ with the Coulomb potential $\phi$. Since 
the time-dependence of the wave functions is a plane wave $\exp(-i \omega t)$,
minimal coupling reduces to the replacement $\omega \rightarrow \omega
- e \phi$. Thus the Dirac equations are
\begin{eqnarray}
	\sqrt{A} \:\alpha^\prime & = & \frac{1}{r} \:\alpha \:-\: 
	((\omega-e \phi) T + m) \:\beta \label{E1} \\
	\sqrt{A} \:\beta^\prime	 & = & ((\omega-e \phi) T - m) \:\alpha \:-\:
	\frac{1}{r} \:\beta \label{E2} \;\;\; .
\end{eqnarray}
In the Einstein equations, we must include the stress-energy
tensor of the electric field \cite{ABS}; this gives
\begin{eqnarray}
	r \:A^\prime & = & 1-A \:-\: 16 \pi (\omega-e \phi) T^2
	\:(\alpha^2 + \beta^2) \:-\: r^2 A T^2\: (\phi^\prime)^2
	\label{E3} \\
	2 r A \:\frac{T^\prime}{T} & = & A-1 \:-\: 16 \pi (\omega-e \phi) T^2 
	\:(\alpha^2+\beta^2) \:+\: 32 \pi \:\frac{1}{r} \:T\: \alpha \beta \:+\:
	16 \pi \:m T \:(\alpha^2-\beta^2) \nonumber \\
&&+ r^2 A T^2\: (\phi^\prime)^2 \;\;\; .
	\label{E4}
\end{eqnarray}
Using current conservation, Maxwell's equations
$\nabla_k F^{jk} = 4 \pi e \sum_{a=1}^2 \overline{\Psi_a} G^j \Psi_a$
reduce to the single second-order differential equation
\begin{equation}
	r^2 A \:\phi^{\prime \prime} \;=\; -8 \pi\: e \:(\alpha^2+\beta^2)
	-  \left( 2r A 	+ r^2 A \:\frac{T^\prime}{T}
	+ \frac{r^2}{2} \:A^\prime \right) \phi^\prime \;\;\; .
	\label{E5}
\end{equation}
The normalization condition for the wave functions \cite[Eqn.\ (5.8)]{FSY} 
remains unchanged, namely
\begin{equation}
	\int_0^\infty (\alpha^2+\beta^2) \:\frac{T}{\sqrt{A}} \:dr \;=\; 
	\frac{1}{4 \pi} \;\;\; .
	\label{C1}
\end{equation}
We seek smooth solutions of these equations, which are asymptotically
Minkowskian and have finite (ADM) mass $\rho$,
\begin{eqnarray}
	\lim_{r \rightarrow \infty} T(r) & = & 1
	\label{C2} \\
	\rho \;:=\; \lim_{r \rightarrow \infty} \frac{r}{2} \:(1-A(r)) &<&
	\infty \label{C3} \;\;\; .
\end{eqnarray}
Furthermore, we demand that the electromagnetic potential vanishes at infinity,
\begin{equation}
	\lim_{r \rightarrow \infty} \phi(r) \;=\; 0 \;\;\; .
	\label{C4}
\end{equation}

The equations (\ref{E1})-(\ref{E5}) are invariant under the gauge 
transformations
\begin{equation}
	\phi(r) \;\rightarrow\; \phi(r) \:+\: \kappa \;\;\;,\spc
	\omega \;\rightarrow\; \omega \:+\: e \kappa \spc ,\kappa \in \R.
	\label{28a}
\end{equation}

\section{Construction of the Solutions}
\setcounter{equation}{0}
The construction of solutions is, (analogous to \cite{FSY}),
simplified by the following scaling argument.
We weaken the conditions (\ref{C1}), (\ref{C2}), (\ref{C4}) to
\begin{equation}
	0 \;\neq\; \int_0^\infty (\alpha^2 + \beta^2) \:\frac{T}{\sqrt{A}}
	\; dr \;<\; \infty \;\;,\;\;\;\;\; 
	0 \;\neq\; \lim_{r \rightarrow \infty} T(r) \;<\; \infty \;\;
	,\;\;\;\;\; \lim_{r \rightarrow \infty} \phi(r) \;<\; \infty \;\;,
	\label{8a}
\end{equation}
and set instead
\begin{equation}
	T(0) \;=\; 1 \;\;\;,\spc \phi(0) \;=\; 0 \;\;\;,\spc m \;=\; 1 \;\;\; .
	\label{1}
\end{equation}
This simplifies the discussion of the equations
near $r=0$; indeed, a Taylor expansion around the origin gives
\begin{eqnarray*}
	\alpha(r) & = & \alpha_1 \:r \:+\: {\cal{O}}(r^2) \;\;\;,\spc
	\beta(r) \;=\; {\cal{O}}(r^2) \\
	A(r) &=& 1 \:+\: {\cal{O}}(r^2) \;\;\;\;\;\;\;\;\!,\spc
	T(r) \;=\; 1 \:+\: {\cal{O}}(r^2) \\
	\phi(r) &=& {\cal{O}}(r^2) \;\;\; .
\end{eqnarray*}
The solutions are now determined by only three real parameters 
$e$, $\omega$, and $\alpha_1$.
For a given value of these parameters, we can construct initial data at
$r=0$ and, using the standard Mathematica ODE solver, we shoot for numerical 
solutions of the modified system (\ref{E1})-(\ref{E5}), (\ref{1}).
By varying $\omega$ (for fixed $e$ and $\alpha_1$), we can arrange that 
the spinors $(\alpha, \beta)$ tend to the origin for large $r$. As one sees 
numerically, the so-obtained solutions satisfy the conditions 
(\ref{C3}) and (\ref{8a}).

For a given solution $(\alpha, \beta, A, T, \phi)$ of this modified system,
we introduce the scaled functions
\begin{eqnarray*}
\tilde{\alpha}(r) & = & \sqrt{\frac{\tau}{\lambda}}
	\:\alpha(\lambda r) \;\;\;,\spc
\tilde{\beta}(r) \;=\; \sqrt{\frac{\tau}{\lambda}}
	\:\beta(\lambda r) \\
\tilde{A}(r) &=& A(\lambda r) \;\;\;\;\;\;\;\;\;\:,\spc
\tilde{T}(r) \;=\; \tau^{-1} \:T(\lambda r) \\
\tilde{\phi}(r) &=& \tau \:\phi(\lambda r) \;\;\; .
\end{eqnarray*}
As one verifies by direct computation, these functions satisfy the original
equations (\ref{E1})-(\ref{C3}) if the physical parameters are
transformed according to
\[ \tilde{m} \;=\; \lambda m \;\;\;,\spc \tilde{\omega} \;=\; 
\lambda \tau\:\omega \;\;\;,\spc \tilde{e} \;=\; \lambda e \;\;\; , \]
where the scaling factors $\lambda$ and $\tau$ are given by
\[ \lambda \;=\; \left( 4 \pi \int_0^\infty (\alpha^2+\beta^2) \:
\frac{T}{\sqrt{A}} \:dr \right)^{\frac{1}{2}} \;\;\;,\spc
\tau \;=\; \lim_{r \rightarrow \infty} T(r) \;\;\; . \]
Then the condition (\ref{C4}) can be fulfilled by a suitable gauge
transformation (\ref{28a}). Notice that $(\tilde{e}/\tilde{m})^2=e^2$ 
is invariant under the scaling. Therefore it is convenient to take 
$(\tilde{e}/\tilde{m})^2$ (and not $\tilde{e}^2$ itself) as the parameter
to describe the strength of the electromagnetic coupling.

We point out that this scaling technique is merely used to simplify
the numerics; for the physical interpretation, however, one must always
work with the scaled tilde solutions.
Since the transformation from the un-tilde to the tilde variables is 
one-to-one, our scaling method yields all the solutions of the original
system. We will from now on consider only the {\em{scaled}}
solutions; for simplicity in notation, the tilde will be omitted.

\section{Properties of the Solutions}
\begin{figure}[tb]
	\epsfxsize=12cm
	\centerline{\epsfbox{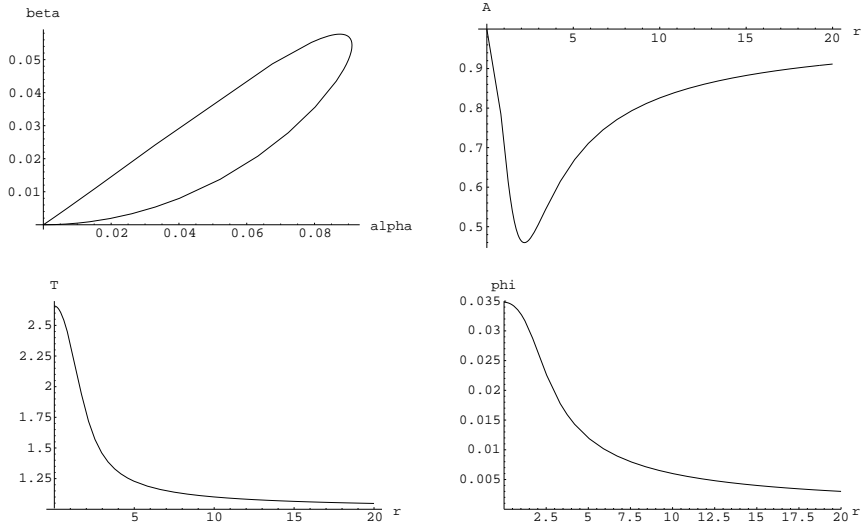}}
	\caption{Solution of the EDM equations for parameter values
	$(e/m)^2=0.7162$, $m=0.7639$, $\omega=0.6807$, $\rho=1.15416$
	($\alpha^\prime(0)=0.05361$).}
	\label{EM_ground}
\end{figure}

\setcounter{equation}{0}
The solutions we found have different rotation number $n=0,1,\ldots$ 
of the vector $(\alpha, \beta)$. In the nonrelativistic limit, $n$ 
coincides with the number of zeros of the corresponding Schr\"odinger 
wave functions, and thus $n=0$ corresponds to the ground state, $n=1$ 
to the first excited state,\ldots, etc. Because of the nonlinearity 
of our equations, $n$ does not in general have this simple 
interpretation. In the following, we will restrict 
ourselves to the $n=0$ solutions. A plot of a typical solution is 
shown in Figure \ref{EM_ground}.

For all considered solutions, the spinors $(\alpha, \beta)$ decay 
exponentially at infinity. This means physically that the fermions
have a high probability of being confined to a neighborhood of the origin.
Since the spinors decay so rapidly at infinity, our solutions 
asymptotically go over into spherically symmetric solutions of the 
Einstein-Maxwell equations; i.e.\ the Reissner-Nordstr\"om solution 
\cite{ABS}. More precisely, the behavior for large $r$ is
\[ A(r) \;\approx\; T(r)^{-2} \;\approx\; 1 \:-\: \frac{2 \rho}{r}
	\:+\: \frac{(2e)^2}{r^2} \;\;\;,\spc
	\phi(r) \;\approx\; \frac{2e}{r} \;\;\; . \]
Thus asymptotically, our solution looks like the 
gravitational and electrostatic field generated by a point particle
at the origin having mass $\rho$ and charge $2e$ .
In contrast to the Reissner-Nordstr\"om solution, however, our 
solutions have no event horizons or singularities. This can be 
understood from the fact that we consider quantum mechanical 
particles (instead of point particles), which implies that the wave 
functions are delocalized according to the Heisenberg Uncertainty 
Principle. As a consequence, the distribution of matter and charge 
are also delocalized, and this prevents the metric from forming 
singularities.

\begin{figure}[tb]
	\epsfxsize=13cm
	\centerline{\epsfbox{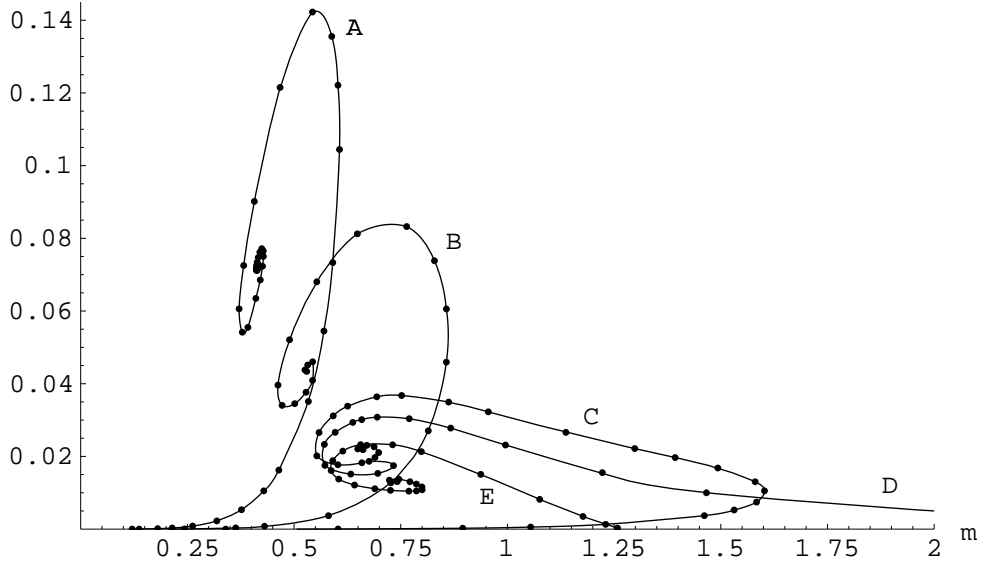}}
	\caption{Binding Energy $m-\omega$ of the Fermions for $(e/m)^2=0$ 
	(A), $0.7162$ (B), $0.9748$ (C), $1$ (D), and $1.0313$ (E).}
	\label{EM_bind}
\end{figure}

The situation near the origin $r=0$, on the other hand, is parametrized
by the rest mass $m$ and the energy $\omega$ of the fermions.
In Figure \ref{EM_bind},
the binding energy $m-\omega$ is plotted versus $m$ for different values 
of the parameter $(e/m)^2$. One sees that $m-\omega$ is always 
positive, which means that the fermions are in a bound state. For 
weak electromagnetic coupling (see Figure \ref{EM_bind}, plots
A and B), the curve has the form of a spiral which starts at 
the origin. The binding energy becomes smaller for fixed $m$ and increasing 
$(e/m)^2$; this is because the electromagnetic repulsion weakens the binding. 
The mass-energy spectrum when $(e/m)^2 \ll 1$ has a similar shape 
as in the case without the electromagnetic interaction \cite{FSY}. The 
stability techniques and results of \cite{FSY} can be generalized directly:
For small $m$, one can use linear perturbation theory to show 
numerically that the solutions are stable (with respect to spherically 
symmetric perturbations). The stability for larger values of $m$ can 
be analyzed with Conley index theory (see \cite{S}), where we take $m$ 
as the bifurcation parameter.
The Conley index theory yields that the stability/instability of a
solution remains unchanged if the parameter m is continuously varied
and no bifurcations take place. Moreover, at the bifurcation points, the
Conley index allows us to analyze the change of stability with powerful
topological methods. We find that all the solutions on the 
``lower branch'' of the spiral (i.e., on the curve from the origin up 
to the maximal value of $m$) are stable, whereas all the solutions on 
the ``upper branch'' are unstable.

The form of the energy spectrum changes when $(e/m)^2 \approx 1$.
This is the regime where the electrostatic and gravitational forces 
balance each other in the classical limit (\ref{newton}). Since the 
assumption of classical point particles does not seem to be appropriate
for our system, we replace (\ref{newton}) by taking the nonrelativistic
limit of the EDM equations more carefully. For this, we assume that $m$
and $e$ are small (for fixed $(e/m)^2$). The coupling becomes weak in this 
limit; thus $A, T \approx 1$, $\phi \approx 0$. The Dirac equations 
give $\omega \approx m$ and $\alpha \gg \beta$.
Therefore the EDM equations go over into the Schr\"odinger equation with the 
Newtonian and Coulomb potentials,
\begin{eqnarray*}
\left( -\frac{1}{2m} \:\Delta \:+\: e \phi \:+\: m V \right) 
\Psi &=& E \:\Psi \\
- \Delta V \;=\; -8 \pi \:m \:|\Psi|^2 \;\;\;,\spc
- \Delta \phi &=& 8 \pi \:e \:|\Psi|^2 
\end{eqnarray*}
(with $E=\omega-m$, $\Psi(r) = \alpha(r)/r$, $V=1-T$; $\Delta$ equals 
the radial Laplacian in $\R^3$).
One sees from these equations that, as in (\ref{newton}), the 
Newtonian and Coulomb potentials are just multiples of each other; 
namely $V=-m \phi/e$. For $(e/m)^2>1$, the total interaction is repulsive,
and the Schr\"odinger equation has no bound states.
We conclude that, in the limit of small $m$, the EDM equations cannot have
particle-like solutions if $(e/m)^2>1$. In other words, the
mass-energy plots of Figure \ref{EM_bind} can only start at $m=0$
if $(e/m)^2<1$. This is confirmed by the numerics (see Figure \ref{EM_bind},
plots C, D, and E). For $(e/m)^2=1$, the plot tends asymptotically to
the point $(m=\infty, \:m-\omega=0)$. It is surprising that we still have
bound states for $(e/m)^2>1$. In this case however, there are only
solutions if $m$ is sufficiently large and smaller than some threshold
value where the binding energy of the fermions becomes zero.

\begin{figure}[tb]
	\epsfxsize=13cm
	\centerline{\epsfbox{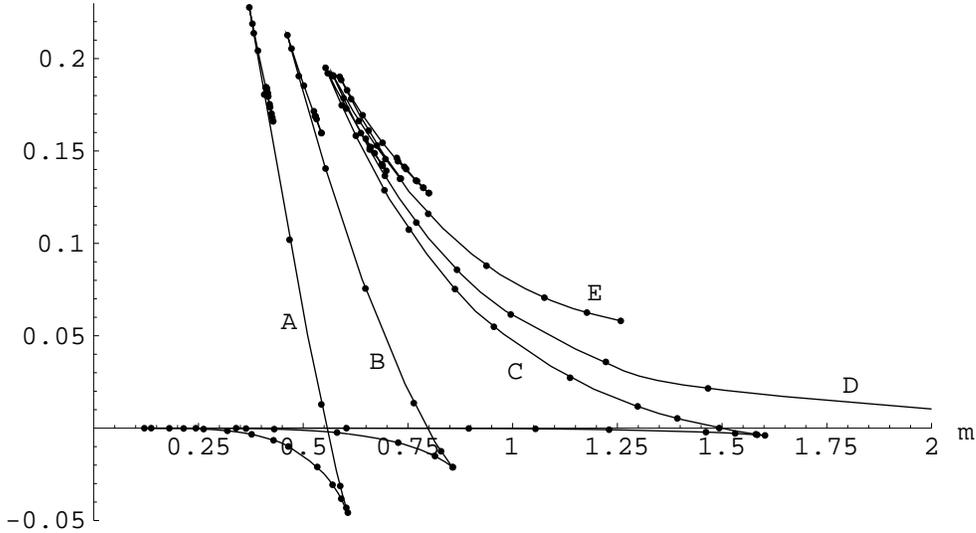}}
	\caption{Total Binding Energy $\rho-2m$ for $(e/m)^2=0$
	(A), $0.7162$ (B), $0.9748$ (C), $1$ (D), and $1.0313$ (E).}
	\label{EM_tm}
\end{figure}

In Figure \ref{EM_tm},
the total binding energy $\rho - 2m$ is plotted 
for different values of $(e/m)^2$. For $(e/m)^2<1$, $\rho - 2m$ 
is negative for the stable solutions. For $(e/m)^2 > 1$, 
however, $\rho - 2m$ is always positive. This indicates that these 
solutions should be unstable, since one gains energy by breaking up 
the binding.

\begin{tabular}{ll}
\\
Mathematics Department, & Mathematics Department,\\
Harvard University, & The University of Michigan,\\
Cambridge, MA 02138  \hspace*{.5cm}(FF \& STY)\hspace*{1cm}
& Ann Arbor, MI 48109 \hspace*{.5cm} (JS)\\
\\
\end{tabular}

email: felix@math.harvard.edu, smoller@umich.edu, yau@math.harvard.edu

\end{document}